\documentclass[pra,twocolumn,eqsecnum,superscriptaddress,showpacs,amsmath,amstex,amssymb,citeautoscript]{revtex4-1}

\usepackage[T1]{fontenc}
\usepackage[utf8]{inputenc}
\usepackage{lipsum}
\usepackage{amsmath}
\usepackage{amssymb}
\usepackage{bbm,bm}
\usepackage{braket}
\usepackage{xcolor}
\usepackage{pifont}
\usepackage[mathscr]{euscript}
\usepackage[shortlabels]{enumitem}
\usepackage[justification=justified]{subcaption}
\usepackage{graphicx}
\usepackage{lipsum}
\allowdisplaybreaks
\usepackage{float}
\usepackage{graphicx}
\usepackage{dsfont}
\usepackage{comment}
\usepackage[colorlinks=true]{hyperref}  
\hypersetup{
    bookmarks=true,         
    unicode=false,          
    pdftoolbar=true,        
    pdfmenubar=true,        
    pdffitwindow=false,     
    pdfstartview={FitH},    
    pdftitle={Displacement-field-tunable superconductivity \\ in an inversion-symmetric twisted van der Waals heterostructure},    
    pdfauthor={Scammell and Scheurer},     
    pdfsubject={},   
    pdfcreator={},   
    pdfproducer={}, 
    pdfkeywords={} {} {}, 
    pdfnewwindow=true,      
    colorlinks=true,       
    linkcolor=blue, 
    citecolor=blue,        
    filecolor=magenta,      
    urlcolor=blue           
}

\renewcommand{\approx}{\simeq}

\definecolor{wrongultramarine}{rgb}{1,0.5,0}

\begin{document}
\title{Anomalous magnetic flux via junction twist-angle in a triplet-superconducting transmon qubit}

\author{Sebastián Domínguez-Calderón}
    \affiliation{School of Engineering and Sciences, Tecnológico de Monterrey, Monterrey 64849, México}

\author{Harley Scammell}
\affiliation{School of Mathematical and Physical Sciences, University of Technology Sydney, Ultimo, NSW 2007, Australia}

\date{\today} 

\begin{abstract}
Superconducting transmon qubits with strong anharmonicity and insensitivity to offset charge are highly desirable for low-error implementation. 
In this work we propose a c-axis junction, comprising triplet superconductors, and set at a relative twist angle. Invoking spin-orbit coupling and spin polarization, which are known to occur in the material platform of choice, we examine the resulting transmon Hamiltonian. This junction allows for direct control of the single and double Cooper pair tunneling strength, and most remarkably, an anomalous magnetic flux---i.e. a phase offset equivalent to magnetic flux, yet in zero magnetic field. Having control over these three parameters---single and double pair tunneling and anomalous flux---allows for optimal design of the transmon qubit. Interestingly, in this architecture, the anomalous flux is determined by the twist angle of the junction, thereby offering a novel zero-field functionality. Our key results rely on symmetry arguments, for concreteness we demonstrate the implementation of our concept using a model of moir\'e graphene-based c-axis junctions. 
\end{abstract}


\maketitle

\section{Introduction} \label{sec:intro}

Superconducting quantum computers have undergone significant improvements and have been utilized to showcase instances of dramatic speedup compared to their classical counterparts \cite{Arute2019}. However, a critical bottleneck in their advancement is errors \cite{KjaergaardReview2020, GyenisPRXQuantum2021b}. As the number of qubits increases, so does the number of error sources, necessitating these errors lie beneath a certain threshold; otherwise, adding more qubits becomes futile. Since true fault-tolerant quantum computation is believed to require formidable hardware resources \cite{FowlerPRA2012,ReiherPNAS2017}, the near-term goal is to dip below the error threshold.  At present, superconducting quantum computers are approaching the error threshold \cite{Acharya2023}, underscoring the crucial importance of further reducing errors in superconducting qubits as a primary avenue of research. 

Recent advancements have pinpointed favorable transmon qubit Hamiltonians that offer tunable anharmonicity and offset charge insensitivity, thereby mitigating errors. These Hamiltonians consist of strong double Cooper pair (CP) tunneling, weak single CP tunneling, and a magnetic flux \cite{patel2023dmon, staples2024universal}. One such architecture is a triple junction {\it d-mon} subject to magnetic flux in one junction, whereby the {\it d-mon} refers to a d-wave/s-wave c-axis superconducting junction. The second is a non-ideal $\pi$-SQUID, as discussed in various works \cite{kitaev2006protected, Paolo_2019, Ciaccia2024, Valentini2024}, but further subjected to a magnetic flux \cite{staples2024universal}.



In contrast to these works, the present study is motivated by two key questions: (i) Since magnetic fields are known inhibitors to achieving small scale devices, can we achieve the desired flux without a magnetic field? (ii) All things being equal, a single-junction qubit is more suited to small scales than a multi-junction qubit; can we design a highly tunable transmon qubit within a single junction?


The present work succeeds in designing a transmon architecture with a high tunability, within a single junction and without magnetic field. Our effective transmon maps onto two of the aforementioned recent architectures: the triple junction {\it d-mon} with magnetic flux \cite{patel2023dmon}, and the (two junction) non-ideal $\pi$-SQUID also subject to a magnetic flux \cite{staples2024universal}. Again we stress that our construction achieves this desirable qubit architecture using all-electrical control (zero magnetic field) and within a single junction. Our key innovation is the use of triplet superconductivity in the presence of spin-orbit coupling (SOC) and intrinsic magnetization. We consider a junction of  spin-triplet superconductors with vectors at some non-zero angle $\Phi_J$ relative to each other on either side of the junction. Then, due to SOC, the twist angle of the junction determines $\Phi_J$. Finally, this $\Phi_J$ acts, via a non-linear relationship, as an {\it anomalous} magnetic flux. This is our key finding. 

Our results are of a general nature, however, for several reasons, we choose to model the key features of moir\'e graphene, e.g. twisted bilayer graphene, because: (i) we require triplet superconductivity, but without nodes in the order parameter, since nodes introduce decoherence \cite{Fominov2003}. A nice way to achieve this is to include an extra quantum number, in our case we choose a valley system which admits a  valley quantum number. (ii) We require a valley system that can host superconductivity and moreover, triplet superconductivity. (iii) We require a superconducting valley system that can simultaneously host magnetic order. (iv) We require a system into which SOC can be induced. These criteria naturally direct us to moir\'e graphene. Notably, for (iv) one can proximity-induce SOC into moir\'e graphene via a transition metal dichalcogenide (TMD) substrate. The form of the induced SOC can be readily tuned based on the relative twist angle graphene and TMD layers \cite{PhysRevB.104.195156,PhysRevB.99.075438,PhysRevB.100.085412,2022arXiv220609478L,PhysRevResearch.4.L022049,PhysRevB.104.075126,PhysRevB.105.115126}.

\section{Background: Triplet superconductivity in a valley system}
We provide a short summary of key details of triplet superconductivity within a valley system, subject to various particle-hole orders (e.g. spin polarization). The particular valley system considered here is modeled on moir\'e graphene.

\begin{figure*}[t]
    \centering
\includegraphics[width=0.85\textwidth]{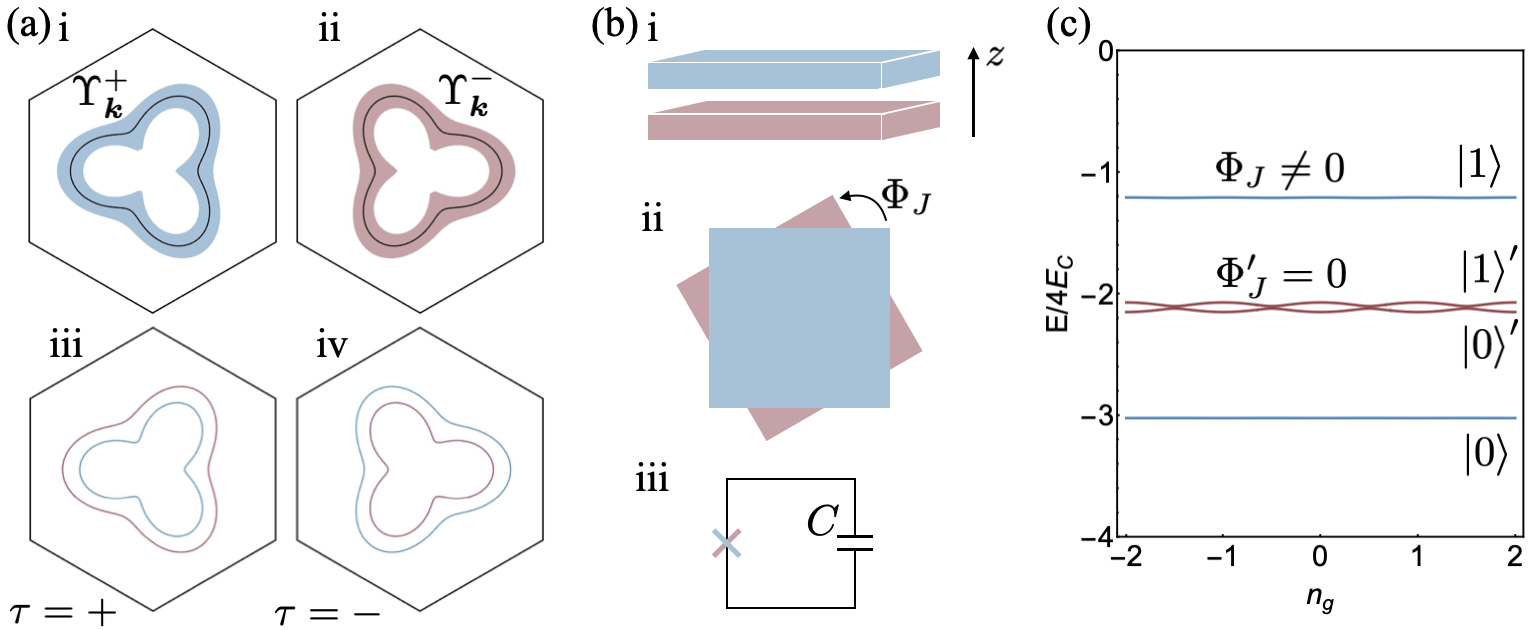}
    \caption{Model details: (a)i\&ii Fermi surfaces of $\varepsilon^0_{\tau\bm k}$ for $\tau=\pm$; gap function $\Upsilon_{\bm k}^\tau$ is superimposed as the shaded region. (a)iii\&iv Fermi surfaces with account of spin-valley locked polarization $\varepsilon_{\tau\bm k}=\varepsilon_{\tau\bm k} + M_{zz} \tau_z s_z$, blue and red correspond to the two spin species. (b)i Schematic of the junction and (b)ii top view of the twisted junction, with twist angle $\Phi_J$. (b)iii Shows the corresponding circuit. (c) Qubit subspace of the transmon spectrum---red eigenvalues $\{\ket{0}',\ket{1}'\}$ corresponds to zero twist angle $\Phi_J'=0$ (denoted with a prime for clarity), blue eigenvalues $\{\ket{0},\ket{1}\}$ corresponds to a non-zero twist angle, $\Phi_J\neq0$.}
    \label{fig1}
\end{figure*}

First, we consider the low-energy moir\'e bands, and to be explicit, we consider doping with chemical potential $\mu>0$. The partially filled moir\'e band is denoted $\varepsilon_{\bm k}$ and is four-fold degenerate due to spin and monolayer-valley degrees of freedom; Pauli matrices acting on these quantum numbers are denoted by $s_\nu$ and $\tau_\nu$, respectively, with $\nu=0, x,y,z$. Moreover, we account for trigonal warping within the dispersion,
\begin{align}
\label{Ek0}
    \varepsilon_{\bm k}=\varepsilon^0_{\bm k}(1 + \lambda \tau_z \cos3\theta_{\bm k})
\end{align}
where $\varepsilon^0_{\bm k}$ is rotationally symmetric, while $\cos3\theta_{\bm k}$ reduces this symmetry to three-fold, where  $\theta_{\bm k}$ is the quasi-momentum angle about a given valley. Here $\lambda<1$ is a dimensionless parameter. See Fig. \ref{fig1}(a) for corresponding Fermi surfaces. 

Next, our model assumes the Cooper pairs are formed from the partially-filled bands, consistent with experimental observations \cite{Park2021_tTLG,HaoKim_tTLG,Jiang-Xiazi_diode,2021arXiv211207127S}. Moreover, we take the Cooper pairs to be {\it extended}-s-wave, i.e. being fully gapped and transforming trivially about the Fermi surface of a given valley. 
These orders have been discussed at various points in the literature \cite{XuBalents2018, You2019, OurClassification, LakeSenthil, DiodeTheory, MobiusPRL, ScammellScheurerDisp2024}. Explicitly, we consider spin-triplet, valley-singlet ordering, with corresponding  gap function
\begin{align}
\label{valleyS}
    {\Delta}_{\boldsymbol{k}}  &=  \Upsilon_{\boldsymbol{k}}
    ({\bm d} \cdot \boldsymbol{{s}}) {s}_y {\tau}_y.
\end{align}
Here $\Upsilon_{\boldsymbol{k}} = \Upsilon_{\boldsymbol{k}}^+\delta_{\tau,+}+\Upsilon_{\boldsymbol{k}}^-\delta_{\tau,-}$ is composed of  two complex scalar functions $\Upsilon_{\boldsymbol{k}}^\tau$ containing the momentum dependence of the superconducting order about the Fermi surfaces (at each valley $\tau$); see the schematic of Fig. \ref{fig1}(a). Meanwhile, the three-component vector ${\bm d}$ accounts for the spin triplet order parameter.

Finally, in addition to superconductivity, we include in our modeling the occurrence of different types of particle-hole orders, such as spin or valley polarization.   
We now provide a toy-model Hamiltonian for the partially filled moir\'e bands of moir\'e graphene with account of particle-hole orders
\begin{align}
\label{hk}
    h_{\bm k}=\varepsilon_{\bm k} -\mu + M_{\mu\nu} \tau_\mu s_\nu
\end{align}
here $M_{\mu\nu}$ account for different possible particle-hole order, e.g. $M_{0z}$ is an out-of-plane spin polarization. A schematic of these key ingredients is provided in Fig. \ref{fig1}(a).

\section{Methods: effective transmon Hamiltonian}
We model a setup of two stacked moir\'e graphene subsystems, separated by a thin insulating layer, and rotated to an angle  $\Phi_J$. An illustration is provided in Fig. \ref{fig1}(b). In this setup, we treat the systems as being two copies of moir\'e graphene, and crucially, the superconducting order from one system is understood to influence the other via the proximity effect. We refer to this as the moir\'e graphene junction --- the relation to a standard Josephson junction will become apparent shortly. 

In the Bogoliubov-de Gennes (BdG) framework, i.e. a mean-field Hamiltonian that captures the electronic normal state as well as the superconducting order parameters, the moir\'e graphene junction is modeled as,
\begin{equation}
\label{Hbdg}
    {H}_{\boldsymbol{k}}  = \frac{1}{2} \begin{bmatrix}
        {h}_{\bm k} & {\Delta}^{(1)}_{\boldsymbol{k}} + e^{i\varphi}{\Delta}^{(2)}_{\boldsymbol{k}} \\
        {\Delta}^{\dagger (1)}_{\boldsymbol{k}} + e^{-i\varphi}{\Delta}^{\dagger (2)}_{\boldsymbol{k}} & -{h}_{-\bm k} \\
    \end{bmatrix},
\end{equation}
where ${\Delta}^{(1)}_{\boldsymbol{k}}$ and ${\Delta}^{(2)}_{\boldsymbol{k}}$ are the gap functions of subsystem 1 and 2, which take the form of \eqref{valleyS}. Note that the magnitude of ${\Delta}^{(2)}_{\boldsymbol{k}}$ will be reduced since it is proximity induced into subsystem 1, see e.g. \cite{patel2023dmon} for an analogous calculation. We have made explicit the U(1) phase difference between these order parameters. We stress now that this phase difference, $\varphi$, and its canonically conjugate angular momentum, ${n}=-i\partial_\varphi$, will be the dynamical phase-space variables of the effective moir\'e graphene transmon Hamiltonian.

To arrive at the transmon Hamiltonian, we compute the free energy to quartic order in the parameter ${\bm d}$, of \eqref{valleyS}. Such an expansion captures both single and double tunneling processes. Explicitly, the (regularized) free energy and subsequent expansion are given by 
\begin{align}
\label{Uexpand}
   \notag {\cal U} &=  \text{Tr}\left[\ln(\omega - {{H}}_{\bm k}) - \ln(\omega - {{H}}^{(0)}_{\bm k}) \right]\\
   \notag &\approx-\frac{1}{2}\text{Tr}\left[{\cal O}_{\bm k}{{\cal G}}_{\omega, \bm k} {\cal O}_{\bm k}{{\cal G}}_{\omega, \bm k}\right] - \frac{1}{4}\text{Tr}\left[\left({\cal O}_{\bm k}{{\cal G}}_{\omega, \bm k} {\cal O}_{\bm k}{{\cal G}}_{\omega, \bm k}\right)^2\right]\\ 
    &\approx a_1 \cos\varphi+b_1 \sin\varphi +a_2 \cos2\varphi+b_2 \sin2\varphi 
\end{align}
here $H^{(0)}_{\bm k}\equiv H_{\bm k}\big|_{\Delta^{(i)}_{\bm k}\to0}$; ${{\cal G}}_{\omega, \bm k} = {G}_{\omega, \bm k} (\eta_0+\eta_z)/2 - {G}_{-\omega, -\bm k}(\eta_0-\eta_z)/2$ and ${G}_{\omega, \bm k} = [i\omega - {h}_{\bm k}]^{-1}$; and ${{\cal O}}_{\bm k} = \eta_+({\Delta}^{(1)}_{\boldsymbol{k}} + e^{i\varphi}{\Delta}^{(2)}_{\boldsymbol{k}}) +\eta_- ({\Delta}^{\dagger (1)}_{\boldsymbol{k}} + e^{-i\varphi}{\Delta}^{\dagger (2)}_{\boldsymbol{k}})$, with $\eta_\mu$ Pauli-matrices accounting for the Nambu space of the BdG Hamiltonian \eqref{Hbdg}. The approximations in passing from the first-to-second and second-to-third lines involves ignoring higher-order tunneling processes, i.e. $\cos n\varphi$ for $n>2$, and ignoring an irrelevant $\varphi$-independent offset. 

Meanwhile, the charging energy of the junction takes the standard form,
\begin{align}
    {\cal K} = 4E_C({n} - n_g)^2,
\end{align}
with ${n}=-i\partial_\varphi$ and $n_g$ an offset {\it gate charge}. 
Together, the moir\'e graphene transmon Hamiltonian is ${\cal H}= {\cal K} + {\cal U}$. Truncating at double CP tunnelling, working in units of $4E_C=1$, and choosing a simpler gauge than \eqref{Uexpand},
\begin{equation}
\label{Htrans}
    {\cal H} = ({n} - n_g)^2 + \alpha_1 \cos(\varphi+\Phi)  +\alpha_2 \cos2\varphi.
\end{equation}
The parameters $\{\alpha_1,\alpha_2, \Phi\}$  control the anharmonicity and charge offset insensitivity (band flatness) of the qubit subspace. For more details of a specific metric, see \cite{patel2023dmon}.  
An ideal regime corresponds to $\alpha_2\gg \alpha_1>1$ with $\Phi\sim\pi/2$ --- a demonstration is provided in Fig. \ref{fig1}(c), which compares the different qubit subspaces formed taking $\{\alpha_1,\alpha_2\} = \{20,2\}$ and with $\Phi=\pi/2$ or $0$. Dramatic improvement, with respect to band flatness (and not shown, also the anharmonicity) can be seen for $\Phi=\pi/2$ (non-zero flux).  The takeaway message is that, the better the control of $\{\alpha_1,\alpha_2, \Phi\}$, the better one can design a transmon qubit. Hence, the objective of the present work is to find setups that allow control of $\{\alpha_1,\alpha_2, \Phi\}$, under the additional constraint of wanting only a single junction and zero magnetic flux. This is to be contrasted with previous proposals, which include the triple junction {\it d-mon} subject to magnetic flux \cite{patel2023dmon} and the non-ideal $\pi$-SQUID: a superconducting loop formed by two $\pi$-periodic circuit elements, also subject to an external magnetic flux \cite{staples2024universal}. 

 \section{Results: Twisted triplet-triplet junction}
As our specific setup, we consider the triplet-triplet junction, i.e. with triplet vectors $\bm d^{(1)}$ and $\bm d^{(2)}$, and further impose an easy axis spin-orbit coupling, via a TMD substrate layer, see e.g. \cite{ScammellScheurerDisp2024} for detailed discussion. The purpose is to reduce the SO(3) spin-rotational symmetry of the $\bm d^{(1,2)}$, and couple it to a given spatial (crystal) axis, i.e. $\bm d^{(1)} = d_0(1,0,0)$ in subsystem 1. Next, spatially rotating subsystem 2 to an angle $\Phi_J$ relative to subsystem 1, implies that $\bm d^{(2)} = d_0(\cos\Phi_J,\sin\Phi_J,0)$; the triplet angle follows the spatial twist angle, thanks to SOC. The total gap function entering the BdG Hamiltonian \eqref{Hbdg} is then 
\begin{align}
{\Delta}^{(1)}_{\boldsymbol{k}} + e^{i\varphi}{\Delta}^{(2)}_{\boldsymbol{k}} = \Upsilon_{\boldsymbol{k}}
    (\bm d^{(1)}+e^{i\varphi}\bm d^{(2)}) \cdot \boldsymbol{{s}} ({s}_y {\tau}_y).
\end{align}
In what follows, we denote $\bm d\equiv \bm d^{(1)}+e^{i\varphi}\bm d^{(2)} $. 

Finally, allowing for spin polarization --- denoted $\bm M = M_{z} {z}$ and taken to be along $z$, i.e. appearing in \eqref{hk} as $M_{z}s_z$ --- the free energy expansion then permits the terms,
\begin{align}
\label{sine_term}
   \delta {\cal U}_1 & \sim i \bm M\cdot (\bm d^*\times \bm d) =  g_1 M_z \sin\Phi_J\sin\varphi,\\
    \notag \delta {\cal U}_2 & \sim i \bm M\cdot (\bm d^*\times \bm d) (\bm d^*\cdot \bm d) = g_2 M_z \sin2\Phi_J\sin2\varphi.
\end{align}
Here $g_{1,2}$ are dimensionless parameters, that can be obtained via direct computation of \eqref{Uexpand}. 
The $\sin\varphi$ and $\sin2\varphi$ contributions to the transmon potential is our key observation. Their existence relies on the vector product $(\bm d^*\times \bm d)\neq\bm0$, i.e. {\it non-unitary} superconductivity \cite{SigristRMP}, as well as the vector $\bm M$ associated here with spin polarization. 

Armed with the above setup, and upon expanding to first-order in $M_z$, we arrive at the parameters of the potential  \eqref{Uexpand} (i.e. before re-gauging to \eqref{Htrans}), 
\begin{align}
\label{VS_params}
 a_1&\sim a_1^0 \cos \Phi_J,\quad
 b_1\sim g_1 M_z \sin\Phi_J,\\
\notag a_2&\sim a_2^0 \cos2\Phi_J,\quad
b_2\sim g_2 M_z \sin2\Phi_J.
\end{align}
Here $a_1^0, a_2^0$ are independent of $M_z$ to order $O(M_z^2)$. Upon re-gauging, we arrive at the transmon Hamiltonian \eqref{Htrans}, with parameters
\begin{align}
\label{Phi_J}
\alpha_1&=\sqrt{a_1^2+b_1^2},\quad \alpha_2=\sqrt{a_2^2+b_2^2},\\
\notag \Phi &= \arctan\left(\frac{g_1 M_z}{a^0_1} \tan\Phi_J\right) - \frac{1}{2}\arctan\left(\frac{g_2 M_z}{a^0_2} \tan2\Phi_J\right).
\end{align}
This constitutes the title claim: that the junction twist angle, $\Phi_J$, generates the anomalous magnetic flux $\Phi$, via a nonlinear relationship \eqref{Phi_J}. 


\

\section{Discussion}
We examined c-axis tunnel junctions built from triplet superconductors. Our objective was to demonstrate that such junctions are useful for creating highly optimal transmon qubits, based on the criteria of anharmonicity and offset charge insensitivity. Leaning on two recent analyses \cite{patel2023dmon, staples2024universal}, we distilled the problem statement down to: how can we arrange the triplet superconductors, so as to have control over the three relevant parameters in the transmon Hamiltonian, $\{\alpha_1,\alpha_2, \Phi\}$. In our work, these parameters are understood to control, respectively, single and double CP tunneling and an anomalous (zero-field) magnetic flux that acts solely on the single tunneling processes. 

To achieve our goal, we found that a spin-triplet junction, with rotated triplet vectors on either side, and in the presence of spin polarization, gave us the desired Hamiltonian. The ability to rotate the spin vector became tied to the ability to rotate the junction, thanks to spin-orbit coupling. 

All of these ingredients: triplet superconductivity, spin polarization, spin-orbit interaction and a valley degree of freedom, are features of moir\'e graphene (with a TMD substrate), which prompted our searches in the first instance. Moreover, taking moir\'e graphene as the qubit platform also gives us hope that the present work will inspire near-term experimental tests, given the ongoing research activity into moir\'e graphene. Aside from moir\'e graphene, we note that triplet superconductivity in a valley system has been modeled via modulated gating of a two-dimensional electron (or hole) gas \cite{LiInghamScammell2020,LiHOTS, scammell2021intrinsic}, which can also be designed to induce spin orbit coupling \cite{ SushkovPRL2013, ScammellPRB2019}. This represents another possible avenue. 

Finally, we contrast with recent works \cite{Diez-Merida2023, deVries2021, Rodan-Legrain2021}, which have achieved gate-defined Josephson junctions within moir\'e graphene (see also recent theory \cite{LawPRR2023}); such junctions are in-plane, whereas we consider the out-of-plane (c-axis) junction. We leave it for future work to explore the ideas presented here within a gate defined, in-plane junction.

\section*{Acknowledgements}
We thank Julian Ingham for critical feedback on the manuscript.

\end{document}